\documentclass{SCGE}
\usepackage[colorlinks,linkcolor=blue,citecolor=blue,urlcolor=blue]{hyperref}
\usepackage{graphicx}
\usepackage{amsmath}
\usepackage{amsfonts}
\usepackage{amssymb}
\usepackage{mathrsfs}
\usepackage{latexsym}
\usepackage{dsfont}
\usepackage{graphicx}

\newcommand{\tr}{{\rm Tr}}

\newcommand{\ket}[1]{|#1\rangle}               %ket
\newcommand{\bra}[1]{\langle #1|}              %bra

\renewcommand{\geq}{\geqslant}
\renewcommand{\leq}{\leqslant}

\newcommand{\dl}{\delta }

\let\citedash\relax
\makeatletter \providecommand{\citedash}{\hbox{-}\penalty\@m}
\makeatother

\begin{document}
%%%%%%新版式要加上这组
%\begin{picture}(0,0){\rm
%\put(0,-20){\makebox[160truemm][l]{\bf {\sanhao\raisebox{2pt}{.}}
%Letter  {\sanhao\raisebox{1.5pt}{.}}}
%}}
%\put(0,-34){\jiuwuhao {\textcolor[rgb]{0.5,0.5,0.5}{\sf %Special Topic: Fluid Mechanics
%}}}%%(11月注释：调\textcolor[rgb]{x,x,x}中的数字x越大越灰)
%\end{picture}

%\input psfig.sty
\def\bm{\boldsymbol}

\def\dl{\displaystyle}
\def\du{\end{document}}
\def\d{{\rm d}}
\def\e{{\rm e}}
\def\i{{\rm i}}

% The author doesn't need fill in it.
%\Year{2016} %
%\Month{January}
%\Vol{59} %  卷号
%\No{1} %  期号
\BeginPage{1} % 起页码
%\AuthorMark{{\rm A. Author}, et al.}  %(11月注释：页眉上的作者)
%\AuthorMarkCite{A. Author, B. Author, C. Author, and D. Author} %(11月注释：citation中的作者)
%\DOI{} % The author doesn't need fill in it.
%\ArtNo{000000}

% \title[short text for running head]{full title}{comments for title}
\title[\xiaosihao Uncertainty Relation Based on Skew Information with Quantum Memory]{Uncertainty Relations Based on Skew Information with Quantum Memory}%标题 {有黑体的时候需要将标题复制在中括号里面，使得引用条显示白体。没有黑体的时候中括号可以删掉}
\author[1]{Zhihao Ma}{}
%\footnote{*Corresponding author (email: dlzhangyu@yahoo.com.cn)}%手动E-mail地址
\author[2]{Zhihua Chen}{}
\author[3,4]{Shao-Ming Fei$^{*}$}{}
\footnote{*Corresponding author (email: feishm@cnu.edu.cn)}
%\author[3]{D. Author}{}

\address[{\rm1}]{School of Mathematical Sciences, Shanghai Jiaotong
University, Shanghai, 200240, China}
\address[{\rm2}]{Department of Applied Mathematics,  Zhejiang
University of Technology, Hangzhou, 310014, China}
\address[{\rm3}]{School of Mathematical Sciences, Capital Normal University, Beijing 100048, China\\
\rm4 Max-Planck-Institute for Mathematics in the Sciences, 04103 Leipzig, Germany}

\maketitle

%\vspace{-3.5mm}
%{\footnotesize\begin{center} Received January 1, 2016; accepted January 1, 2016; published online January 1, 2016%收稿日期
%\end{center}}
%\vspace*{-5mm}

%     Abstract is required.
%\begin{center}
%\rule{16.5cm}{0.4pt}
%\parbox{16.5cm}
%{\begin{abstract}We present a new uncertainty relation by defining a measure of uncertainty based on skew information.
%For bipartite systems, we establish uncertainty relations with the existence of a quantum memory.
%A general relation between quantum correlations and tight bounds of uncertainty has been presented.%摘要
%\end{abstract}}
%\end{center}\vspace*{-0.6cm}

%\begin{center}
%\parbox{16.5cm}
%{\bf\jiuhao Uncertainty relation, Skew information, Quantum correlation}%关键词
%\end{center}

\begin{center}
{}%分类号(3~5 codes)  http://phys.scichina.com:8083/sciGe/UserFiles/File/pacs.pdf
%\CITA    %%(11月注释：Citation内容自动生成)
%\Cit{~~~???, et al. ???. Sci China-Phys Mech Astron, 2014, 57: 1--6, doi:}%%(11月注释：Citation内容需手动填写)
\end{center}

%\textwidth=178truemm \textheight=236truemm%%%%%%新版式要加上

%%%%%%%%%%%%%%%%%%%%%%%%%%%%%%%%%%%%%%%%%%%%%%%%%%%%%%%%%%%%
%\wuhao\vspace*{1.5mm}

\begin{multicols}{2}

%%%%%%%%%%%%%%%%%%%%%%%%%%%%%%%%%%%%%%%%%%%%%%%%%%%%%%%%%%%%
%% Text of article.
%%%%%%%%%%%%%%%%%%%%%%%%%%%%%%%%%%%%%%%%%%%%%%%%%%%%%%%%%%%%
%    Section headings
\renewcommand{\baselinestretch}{1.08} \baselineskip 12.2pt\parindent=10.8pt

\renewcommand{\thefootnote}

\noindent Dear Editors,

\noindent Uncertainty principle is one of the most fascinating features of the quantum world.
It asserts a fundamental limit on the precision with which
certain pairs of physical properties of a particle, such
as position and momentum, can not be simultaneously known.
The uncertainty principle has attracted considerable attention
since the innovation of quantum mechanics and has been investigated in terms of various types of uncertainty inequalities:
%in terms of the noise and disturbance, according to successive measurements,
as informational recourses in
entropic terms, by means of majorization technique
and based on sum of variances and standard deviations.

For a pair of observables $R$ and $S$, the well-known Heisenberg-Robertson uncertainty relation \cite{Heisenberg} says that, $V_{\rho}(R).V_{\rho}(S)\geq \frac{1}{4}|\tr{\rho\,[R,S]}|^{2},$
where  $[R,S]=RS-SR$ is the commutator and $V_{\rho}(R)$ is the standard deviation of $R$.
The entropies serve as appropriate measures of the information content.
It is also used to quantify the quantum uncertainty:
the sum of the Shannon entropy of the probability distribution of the  outcomes  is no less than $\log_2\frac{1}{c}$ \cite{y11} when $R$ and $S$ are measured.
%$c=\max_{j,k}|\langle \psi_j|\phi_k\rangle|^{2}$, $\ket{\psi_j}$ and $\ket{\phi_k}$ are the eigenvectors of observables $R$ and $S$, respectively.
The term $1/c$ quantifies the complementarity of the two observables $R$ and $S$. It has been proved that the entropy uncertainty relations do imply the Heisenberg's uncertainty relation.

Concerning bipartite systems,
%the uncertainty relations become more interesting and may depend on
%the correlations between the subsystems in general.
the authors in \cite{Berta} provided a bound on the uncertainties which depends on the amount of entanglement between the measured particle $A$ and the quantum memory $B$.
%the sum of the uncertainty about the outcome of measurement $R$ and $S$ both given information stored in a quantum memory B,
%is lower bounded by
%for any bipartite density matrix $\rho_{AB}$ in tensor space $H_{A}\otimes H_{B}$, the following uncertainty relation holds:
%\begin{equation}
%S(R|B)+S(S|B)\geq \log_{2}\frac{1}{c}+S(A|B)\label{bi-entropy}
%\end{equation}
%where $S(R|B)$, $S(S|B)$ and $S(A|B)$ are the conditional entropies.
%$H(R|B)$ is used to represent the uncertainty about the outcome of measurement $R$ given information stored in a quantum memory $B$ . $H(A|B)$ quantifies the %amount of entanglement between the particle and the memory.
The result of \cite{Berta} was further improved  to depend on the quantum discord between particles $A$ and $B$ in \cite{Pati}.
Recently the authors in\cite{Fan} obtained  entropic uncertainty relations for multiple measurements with quantum memory.
%\begin{eqnarray}
%S(R|B)+S(S|B)\geq \log_{2}\frac{1}{c}+S(A|B)\nonumber\\
%\hspace{0.5cm}+\max\{0, D_{A}(\rho_{AB})- C_{A}(\rho_{AB})\},
%\end{eqnarray}
%where $D_{A}(\rho_{AB})$ is the quantum discord, $C_{A}(\rho_{AB})$ is the classical correlation
%with respect to the measurement on subsystem A.

The quantum uncertainty relation can be also described in terms of skew information.
$I\left(\rho,H\right) =-\frac{1}{2}\tr\left[\sqrt{\rho},H\right]^{2}$ is introduced to quantify the degree
of non-commutativity of a state $\rho$ and an observable $H$,
which is reduced to the variance $V_{\rho}(H)$ when $\rho$ is a pure state \cite{wigner}.
It can be interpreted as quantum uncertainty of $H$ in $\rho$.
Luo introduced another quantity in \cite{luo}, $J_{\rho}(H)=\frac{1}{2}\tr[(\{\sqrt{\rho},H_{0}\})^{2}]$,
where $\{X,Y\}$ is the anti-commutator, $H_{0}=H-\tr(\rho H)I$ with $I$ the identity operator.
The following inequality holds \cite{luo},
\begin{equation}
\sqrt{I\left(\rho,R\right)J_{\rho}(R)}\sqrt{I\left(\rho,S\right)J_{\rho}(S)}\geq \frac{1}{4}|\tr(\rho[R,S])|^{2}.\label{LUO}
\end{equation}
%Inequality (\ref{LUO}) can be also rewritten as
%$\sqrt{I\left(\rho,R\right)I\left(\rho,S\right)}\geq L_{\rho}(R,S)$,
%where $L_{\rho}(R,S)$ is defined by $L_{\rho}(R,S):=\frac{1}{4}\frac{|\tr(\rho[R,S])|^{2}}{\sqrt{J_{\rho}(R)J_{\rho}(S)}}$.
%Here when $J_{\rho}(R)J_{\rho}(S)=0$, $L_{\rho}(R,S)$ is defined to be zero.
$\sqrt{I\left(\rho,R\right)J_{\rho}(R)}$ can be regarded as a kind of measure for quantum uncertainty. Hence we define $\mathcal{UN}(\rho, R):=\sqrt{I\left(\rho,R\right)J_{\rho}(R)}$. Then we define the uncertainty of $\rho$ associated to the projective measurement $\{\phi_k\}$ as:
$\mathcal{UN}\left(\rho\right)_{\{\phi_k\}}=\sum_{k}\mathcal{UN}\left(\rho,\phi_{k}\right)=\sum_{k}\sqrt{I\left(\rho,\phi_{k}\right)J_{\rho}(\phi_{k})},$
where $\phi_k=\ket{\phi_k}\bra{\phi_k}$ and $\psi_k=\ket{\psi_k}\bra{\psi_k}$ are
the rank one spectral projectors of two non-degenerate observables $R$ and $S$ with eigenvectors $\ket{\phi_k}$ and $\ket{\psi_k}$, respectively.

Now we consider the case of bipartite state $\rho_{AB}$ in tensor space $H_{A}\otimes H_{B}$\cite{Uhlmann}. Recall that quantum discord
%\cite{Ollivier,Henderson},
is a kind of quantum correlation that is different from the entanglement and has found many novel applications\cite{Modi}.
%Quantum discord was defined as the minimal difference of mutual information, before and after local projective measurement on $H_{A}$, and
A bipartite state $\rho_{AB}$ is of zero discord if and only if it is a classical-quantum correlated state (CQ state).
%$\rho^{AB}=\sum\limits_{k} \lambda_{k} \phi_k \otimes \rho^{B}_{k}$.
Besides the definition of the original discord, there are some other discord-like measures sharing the same properties such that their values are zero iff the state is a CQ state.
In this letter we define another discord-like measure. Let $\mathcal{O}$ be any orthogonal basis space in Hilbert space $H_{A}$. Let $\ket{\phi_k}$ be an orthogonal basis of $H_{A}$ and
$\phi_k=\ket{\phi_k}\bra{\phi_k}$ the orthogonal projections on $H_{A}$.
We define the quantum correlation of $\rho_{AB}$ as:
\begin{eqnarray}
&&\mathcal{Q}\left( \rho^{AB}\right)
=\min_{\mathcal{O}}\sum\limits_{k}[I(\rho^{AB},\phi_{k}\otimes I_{B})- I(\rho^{A},\phi_k)],
\end{eqnarray}
where the minimum is taken over all the orthogonal bases in $H_{A}$, $\rho^{A}$ is the reduced state of system $A$.

From the inequality in \cite{Lieb} that $I(\rho^{AB},X\otimes I)\geq I(\rho_{A},X)$
for any bipartite state $\rho^{AB}$ and any observable $X$ on $H^{A}$,
we have the property that $\mathcal{Q}\left( \rho^{AB}\right)\geq 0$.
Moreover, $\mathcal{Q}\left( \rho^{AB}\right)= 0$ if and only if $\rho_{AB}$ is a CQ state  by using the method in proving the theorem 1, property (1) of \cite{luo12}.

%From \cite{Lieb}, one has that for any bipartite state $\rho^{AB}$ and any observable $X$ on $H^{A}$,
%$I(\rho^{AB},X\otimes I)\geq I(\rho_{A},X)$.
%It follows that for any set of rank one orthogonal projections  $\{\phi_k\}$ on $H_{A}$:
%$\sum\limits_{k}[I(\rho_{AB},\phi_k\otimes I^{B})- I(\rho_{A},\phi_k)]\geq 0$.
%Therefore we have $\mathcal{Q}\left( \rho^{AB}\right)\geq 0$.
%Moreover, $\mathcal{Q}\left( \rho^{AB}\right)= 0$ if and only if $\rho_{AB}$ is
%a CQ state, so it is a discord-like measure. This can be seen easily .

%The definition of the quantum correlation $\mathcal{Q}\left( \rho^{AB}\right)$
%looks similar to the quantum correlation measures introduced

%However, they are quite different.
$\mathcal{Q}\left( \rho^{AB}\right)$ has a term of measurement on the subsystem $A$, which gives an explicit physical meaning: it is the minimal difference of incompatibility of
the projective measurements on the bipartite state $\rho^{AB}$ and on the local reduced state $\rho^{A}$.
It quantifies the quantum correlations between the subsystems $A$ and $B$.

{\bf Theorem} Let $\rho^{AB}$ be a quantum state on  $H_{A}\otimes H_{B}$, $\{\phi_k\}$ and $\{\psi_k\}$
denote two sets of rank one projective measurements on $H^{A}$. Then the following uncertainty relation holds:
\begin{eqnarray}
&&\mathcal{UN}\left(\rho^{AB}\right)_{\{\phi_k\otimes I^{B}\}}+\mathcal{UN}\left(\rho^{AB}\right)_{\{\psi_k\otimes I^{B}\}}
\nonumber\\&&
\geq \sum_{k}2L_{\rho^{A}}(\phi_{k},\psi_{k})+2\mathcal{Q}\left( \rho^{AB}\right).\label{un-a}
\end{eqnarray}
where $L_{\rho^{A}}(\phi_{k},\psi_{k})=\frac{1}{4}|\tr(\rho^{A}[\phi_{k},\psi_{k}])|^{2}/
\sqrt{J_{\rho^{A}}(\phi_{k})J_{\rho^{A}}(\psi_{k})}$.

{\bf Proof.} By definition we have
\begin{eqnarray}
&&\mathcal{UN}\left(\rho^{AB}\right)_{\{\phi_k\otimes I^{B}\}}+\mathcal{UN}\left(\rho^{AB}\right)_{\{\psi_k\otimes I^{B}\}}\nonumber\\
&&\geq\sum_{k}I\left(\rho^{AB},\phi_{k}\otimes I^{B}\right)+\sum_{k}I\left(\rho^{AB},\psi_{k}\otimes I^{B}\right)\nonumber\\
&&=\sum_{k}I\left(\rho^{A},\phi_{k}\right)+\sum_{k}I\left(\rho^{A},\psi_{k}\right)\nonumber+\sum_{k}[I\left(\rho^{AB},\phi_{k}\otimes I^{B}\right)\nonumber\\
&&-I\left(\rho^{A},\phi_{k}\right)]+\sum_{k}[I\left(\rho^{AB},\psi_{k}\otimes I^{B}\right)-I\left(\rho^{A},\psi_{k}\right)]\nonumber\\
&&\geq\sum_{k}2\sqrt{I\left(\rho^{A},\phi_{k}\right)I\left(\rho^{A},\psi_{k}\right)}+\sum_{k}[I\left(\rho^{AB},\phi_{k}\otimes I^{B}\right)\nonumber\\
&&-I\left(\rho^{A},\phi_{k}\right)]+\sum_{k}[I\left(\rho^{AB},\psi_{k}\otimes I^{B}\right)-I\left(\rho^{A},\psi_{k}\right)]\nonumber\\
&&\geq \sum_{k}2L_{\rho^{A}}(\phi_{k},\psi_{k})+2\mathcal{Q}\left( \rho^{AB}\right).\ \ \ \label{un-b}
\end{eqnarray}
The first inequality holds since $I\left(\rho,R\right)\leq J_{\rho}(R)$ (see \cite{Furuichi}). The second inequality is due to the Cauchy-Schwarz inequality. The final inequality holds because the optimal measurement for $\mathcal{Q}\left( \rho^{AB}\right)$
may not be $\phi_{k}$ or $\psi_{k}$.

From the theorem, we obtain an uncertainty relation in the form of sum of skew information, which is in some sense similar to the result in the recent work \cite{chenbin}. But actually our result is quite different from that in \cite{chenbin}, which only deals with single partite case.
We treat the bipartite case with a quantum memory $B$. Interestingly, the lower bound contains two terms, one term is the quantum correlation $\mathcal{Q}\left( \rho^{AB}\right)$, the other term is $\sum_{k}L_{\rho^{A}}(\phi_{k},\psi_{k})$ which characterizes the degree of complementarity of two measurements, just as the meaning of $\log_{2}\frac{1}{c}$ in the entropic uncertainty relation \cite{Berta}. Therefore our result can be viewed as an analogue of the bipartite entropic uncertainty relation.

As an example, we consider the 2-qubit Werner state $\rho=\frac{2-p}{6}I_4+\frac{2p-1}{6}V,$
where $p\in[-1,1]$ and $V=\sum\limits_{kl}|kl\rangle\langle kl|$.
Take $\sigma_x$ and $\sigma_z$ as the two observables.
For our theorem we have the values of the left hand side and the right hand side of (\ref{un-a}), $\frac{\sqrt{5-2\sqrt{3-3p^2}-2p(1-p+\sqrt{3-3p^2})}}{3}$ and
 $\frac{2-p-\sqrt{3-3p^2}}{3}$, respectively.
While the left hand side of the theorem in \cite{luo} is $\frac{1}{9}(2-p-\sqrt{3-3p^2})(4+p+\sqrt{3-3p^2})$ and the right hand side is 0.
From the result in \cite{Pati}, the left hand side is $-\frac{2(2-p)}{3}\log_2\frac{2-p}{6}-\frac{2(1+p)}{3}\log_2\frac{1+p}{6}-2$,
the  bound is the same as the left hand side, see Fig. 1 for comparision.

\begin{figure}[H]%"[]"中为位置参数，四个参数tbph依次是置顶、置底、浮动、当前位置，，选用的参数优先顺序为h-t-b-p
\centering
\includegraphics[width=0.9\columnwidth]{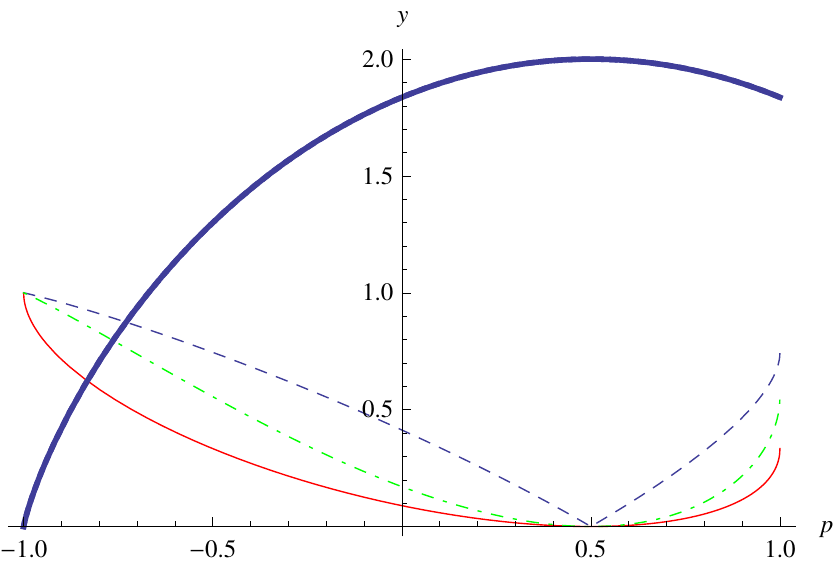}%"scale"后的数字为图形的宽度，也可用"width=1.0\columnwidth" 定义
\caption{Axe y is the uncertainty and its lower bound. The result in \cite{Pati} is represented by blue thick line, the corresponding lower bound is represented by blue dotted line; the result in \cite{luo} is represented by green dotdashed line, the lower bound is 0; the left hand side of
our theorem is represented by blue dashed line and the lower bound of our theorem  is represented by red line.} %图题
\label{fig:example5}%{}中"fig:example1"为图名，引用时用\ref{fig:example1}
\end{figure}

In summary, we have established a new uncertainty relation based on the skew information. We studied the case of uncertainty relation with the existence of a quantum memory for the bipartite quantum system. Our result shows that quantum correlations can be used to obtain a tighter bound of uncertainty.

\Acknowledgements{\bahao Zhihao Ma and Zhihua Chen thank Davide Girolami for useful discussions.
This work is supported by NSFC under numbers 11275131, 11371247, 11571313 and 11675113.}

\end{multicols}

\end{document}